# Intertwined magnetism and charge density wave order in kagome FeGe

Xiaokun Teng[1*], Ji Seop Oh[1,2*], Hengxin Tan[3], Lebing Chen[1], Jianwei Huang[1], Bin Gao[1], Jia-Xin Yin[4], Jiun-Haw Chu[5], Makoto Hashimoto[6], Donghui Lu[6], Chris Jozwiak[7], Aaron Bostwick[7], Eli Rotenberg[7], Garrett E. Granroth[8], Binghai Yan[3], Robert J. Birgeneau[2,9#], Pengcheng Dai[1#], Ming Yi[1#]

[1]Department of Physics and Astronomy, Rice University, Houston, Texas, USA

[2]Department of Physics, University of California, Berkeley, CA, USA

[3]Department of Condensed Matter Physics, Weizmann Institute of Science, Rehovot, Israel

[4]Laboratory for Topological Quantum Matter and Advanced Spectroscopy, Department of Physics, Princeton University, Princeton, NJ, USA

[5]Department of Physics, University of Washington, Seattle, WA, USA

[6]Stanford Synchrotron Radiation Light Source, SLAC National Accelerator Laboratory, Menlo Park, CA, USA

[7]Advanced Light Source, Lawrence Berkeley National Laboratory, Berkeley, CA, USA

[8]Neutron Scattering Division, Oak Ridge National Laboratory, Oak Ridge, TN, USA

[9]Materials Science Division, Lawrence Berkeley National Laboratory, Berkeley, CA, USA

*These authors contributed equally*

#Corresponding authors, email: robertjb@berkeley.edu, pdai@rice.edu, mingyi@rice.edu

**Electron correlations often lead to emergent orders in quantum materials[1]. Kagome lattice materials are emerging as an exciting platform for realizing quantum topology in the presence of electron correlations[2]. This proposal stems from the key signatures of electronic structures associated with its lattice geometry: flat band induced by destructive interference of the electronic wavefunctions, topological Dirac crossing, and a pair of van Hove singularities (vHSs)[3-6]. A plethora of correlated electronic phases have been discovered amongst kagome lattice materials, including magnetism, charge density wave (CDW), nematicity, and superconductivity[7-19]. These materials can be largely organized into two types: those that host magnetism[7-14,15,16] and those that host CDW order[17,18,19]. Recently, a CDW order has been discovered in the magnetic kagome FeGe[20], providing a new platform for understanding the interplay between CDW and magnetism. Here, utilizing angle-resolved photoemission spectroscopy, we observe all three types of electronic signatures of the kagome lattice: flat bands, Dirac crossings, and vHSs. From both the observation of a temperature-dependent shift of the vHSs towards the Fermi level as well as guidance via first-principle calculations, we identify the presence of the vHSs near the Fermi level ($E_F$) to be driven by the development of underlying magnetic exchange splitting. Furthermore, we show spectral evidence for the CDW order as gaps that open on the near-$E_F$ vHS bands, as well as evidence of electron-phonon coupling from a kink on the vHS band together with phonon hardening observed by inelastic neutron scattering. Our observation points to the magnetic interaction-driven band modification resulting in the formation of the CDW order, indicating an intertwined connection between the emergent magnetism and vHS charge order in this moderately-correlated kagome metal.**

In quantum materials where the energy scale of electron-electron correlations is comparable to that of the electronic kinetic energy, a range of emergent electronic phases are found. Well-known systems that exhibit such rich phase diagrams include the unconventional superconductor families of cuprates[21] and iron-based superconductors[22-24], where magnetic order, charge density wave order, nematicity, and superconductivity have been found in close proximity due to the entwinement of a multitude of degrees of freedom with similar energy scales, including spin, charge, and orbitals. Besides engendering large quantum fluctuations that may give rise to novel phases, the intertwinement of multiple orders allows tunability of the ground state via the coupled order parameters. It is expected that when quantum topology is realized in such a strongly correlated regime, unprecedented exotic phases are to be discovered. Recently, kagome lattices have been extensively investigated for the rich emergent physics associated with its lattice geometry in the presence of topology[2-20]. Amongst the insulators, quantum spin liquid states may exist due to the magnetic frustration[2,25]. Analogously, the geometric frustration for an electronic state could also cause localization as a way to quench the kinetic energy, thereby leading to a regime with narrow electronic bandwidth where electron correlation effects could be strong enough to induce emergent phases[3-6]. Equally importantly, the point group of the kagome lattice is that of graphene, and hence the kagome lattice also shares many properties of the Dirac fermions. Therefore, kagome lattices have been intensely explored as a model system to realize topology in the presence of strong electron correlations.

Amongst the known kagome metals, some exhibit magnetic order with ferromagnetically ordered sheets that are either ferromagnetically or antiferromagnetically stacked[9-11,13,14,15,16,23], with ordering temperatures often exceeding room temperature. Another class of recently discovered kagome materials, AV$_3$Sb$_5$ (A = K, Rb, Cs)[17-19], hosts CDW order near 100 K that may be associated with a time-reversal symmetry breaking field[27,28]. The interpretation on the origin of these orders ranges from those initiated from a strongly interacting picture to those based on electronic instabilities from band structure[29-33]. For a two-dimensional (2D) kagome lattice, three key features have been identified in the electronic structure (Fig. 1d): i) a flat band derived from the destructive interference; ii) a Dirac crossing at the Brillouin zone (BZ) corner (K point in the inset of Fig. 1d); and iii) a pair of vHSs at the BZ boundary (M point). It has been proposed that the large density of states from the kagome flat bands could induce ferromagnetism[5,6]. Alternatively, at special fillings where the vHS is at $E_F$, interaction between the saddle points could also induce CDW order[4,31,32]. The large energy separation between the flat band and the vHSs (Figs. 1d and 1e) may be one reason why CDW order and magnetic order have not been commonly observed simultaneously within one system. Efforts at tuning the ordered phases include tuning the energy location of either the flat bands or vHSs relative to the chemical potential via chemical doping[34], hydrostatic pressure[35], and uniaxial strain[36], all of which have demonstrated potential, yet so far have resulted in decoupled order parameters of either magnetism or CDW order.

Very recently, it has been discovered that a CDW order appears deep in a magnetically ordered kagome metal—FeGe[20,37], providing the opportunity to explore the possible connection between magnetism and CDW order on a kagome lattice. Hexagonal FeGe[38-40] (Figs. 1a-b), isostructural to FeSn and CoSn, consists of stacks of Fe kagome planes with both in-plane and inter-plane Ge atoms. A sequence of phase transitions has been found (Fig. 1c). First, an A-type antiferromagnetic (AFM) order appears below $T_N \approx 410$ K, with moments aligned ferromagnetically (FM) along the c-axis within each plane and anti-aligned between layers (Fig.

1a). At 100 K where CDW order takes place ($T_{CDW}$), two types of CDW order were found with $\mathbf{Q}_{CDW1}$ = (0.5, 0, 0)/(0, 0.5, 0) that is completely in-plane and $\mathbf{Q}_{CDW2}$ = (0.5, 0, 0.5)/(0, 0.5, 0.5) that also has an out-of-plane component, similar to that found in the AV$_3$Sb$_5$ systems[18,19]. Finally, at a lower temperature $T_{Canting} \approx 60$ K, the magnetic moments cant from the $c$-direction to give a $c$-axis double cone (canted) AFM structure. In addition, evidence of strong coupling between the CDW order and AFM order has been reported in the form of an enhanced magnetic moment at the onset of the CDW order[20].

Here we use a combination of high resolution angle-resolved photoemission spectroscopy (ARPES), inelastic neutron scattering (INS), and density functional theory (DFT) to study the formation of the CDW order from the magnetic phase to gain insights into the potential contributors to the origin of CDW order in magnetic FeGe. In particular, in the AFM state, we observe saddle points in the vicinity of $E_F$, as well as Kagome flat bands and Dirac cones. With the guidance of DFT calculations, we identify an orbital-selective contribution of vHS to a potential nesting condition near $E_F$, enabled by the spin-splitting of the bands within each ferromagnetic Kagome layer. At $T_{CDW}$, gaps open on the vHS bands across the $\mathbf{Q}_{CDW}$ vectors. In addition, we find that the electronic states on the vHS bands couple to an optical phonon mode. We therefore identify the key ingredients to the formation of CDW in Kagome FeGe to be the presence of orbital-selective vHS near $E_F$ induced by the spin splitting of the ordered moment as well as electron-phonon coupling. The observations taken together suggest an intimate connection between the CDW order and magnetism on the kagome lattice and lay the groundwork for theoretical efforts incorporating electron correlations to identify CDW order formation in a magnetic kagome lattice.

We begin with an overview of the electronic structure of FeGe from DFT calculations. Consistent with previous first-principle calculations[41,42], the density of states (DOS) calculated for the paramagnetic (PM) phase exhibit a large peak at $E_F$, which splits into two in the AFM phase (Fig. 1f). By directly comparing to the band structures calculated in each phase (Fig. 1e), the peaks in the DOS are indeed associated with the kagome flat bands. Moreover, despite that the magnetic structure in FeGe is AFM, each kagome layer is ferromagnetic with a well-defined spin-majority and spin-minority contribution. From Fig. 1e, we confirm that the flat bands above $E_F$ in the AFM phase correspond to the spin minority bands while those below $E_F$ correspond to the spin majority bands within each FM layer. Hence one can largely interpret the AFM band structures as an interaction-driven magnetic splitting of the kagome bands within each FM layer (Fig. 1d). Interestingly, the consequence of such splitting is that the vHSs that were relatively far from $E_F$ in the PM state are brought to the vicinity of $E_F$ in the AFM phase in the spin minority sector, as shown by the marked vHS1 and vHS2 in Fig. 1e. Although DFT cannot capture all the details of the experimental observations because this is a moderately-correlated system, we will show in the following that this qualitative understanding is consistent with our experimental observations.

Next, we gauge the strength of the electron correlations in FeGe by comparing the measured DOS with DFT. From our angle-integrated photoemission spectrum, we can identify key spectral features that match with that from the calculated DOS in the AFM phase (see Supplementary Information (SI) Fig. S1), which allows us to extract an overall renormalization factor of 1.6, close to that in some of the iron-based superconductors[43]. With such a renormalization factor, the DFT calculated band structure for the AFM state can provide a reasonable match with that of the measured dispersions covered in a large energy range (Fig. 2c). In particular, we can identify a broad band near -0.75 eV that matches the location of the spin-majority kagome flat bands. Next, we examine the key signatures of the kagome electronic structure near $E_F$ in the AFM phase. From a detailed photon energy measurement, we identify 69 eV photons to cross the $k_z = 0$ plane while 47 and 102 eV photons to cross the $k_z = \pi$ plane (see Figure S3). Since the photoemission process is subject to substantial $k_z$-broadening effect[44], we discuss the following using the 2D projected notation of the BZ (Fig. 2a). We find two types of terminations amongst the samples measured, corresponding to the Ge and Fe-kagome terminations, similar to the case of FeSn[10]. From a detailed analysis of the two terminations (see Figures S2, S3, S4, S6), we can identify two sets of vHSs and Dirac cones (DCs) that are termination-independent and therefore intrinsic to the bulk kagome layers. The first set (marked in blue as vHS1 and DC1) consists of linear bands that cross at –0.6 eV at the BZ corner, $\bar{K}$ points (Fig. 2d). The Dirac cone structures can be better visualized in a stack of constant energy contour plots in Fig. 2e, where circular contours shrink down to the Dirac point and expand out again. Furthermore, the linear dispersions from DC1 rise to cross $E_F$, forming a vHS on the BZ edge $\bar{M}$ point (Fig. 2d-e). The saddle point nature of this vHS can be better visualized from a series of cuts across the $\bar{M}$ point (Fig. 2f), where the band top of the hole-like band parallel to $\bar{K}$-$\bar{M}$-$\bar{K}$ has a minimum energy at the $\bar{M}$ point, indicative of the electron-like nature in the orthogonal direction. The Fermi momenta ($k_F$) of this vHS1 (denoted by blue dots) form circular Fermi pockets (Fig. 2e). Besides this set of DC and vHS, a second set (labeled in dark green) consists of a DC2 at -0.2 eV that is connected to vHS2 slightly below $E_F$ with a much smaller band velocity compared to that of vHS1. In addition, we also identify a DC3 at -0.06 eV. For the hexagonal lattice symmetry, the five Fe $3d$ orbitals are split into three groups, $d_{x^2-y^2}/d_{xy}$, $d_{xz}/d_{yz}$, and $d_{z^2}$. Guided by orbital-projected DFT calculations and consideration of both in-plane and out-of-plane band velocities (see Figure S4, S5), we identify DC1/vHS1 to be of $d_{x^2-y^2}/d_{xy}$ origin, DC2/vHS2 to be of $d_{xz}/d_{yz}$ origin, and DC3 to be of $d_{z^2}$ origin. A summary of the identified vHSs and DCs is shown in Fig. 2b. We note that while common features can be identified between DFT and measured dispersions, the precise locations of the vHSs are not well captured by DFT even with the overall renormalization, suggesting non-negligible correlation effects at play.

If the qualitative understanding of the spin-splitting from DFT calculations is correct, the vHSs near $E_F$ observed in the AFM state must be the spin minority set that is shifted up as the magnetic moment orders with lowering temperature. To check for this, we examine the temperature evolution of vHS1, which as a mostly in-plane orbital can be tracked more accurately as it is less affected by $k_z$-broadening compared to vHS2. In Figs. 3e-g, we show the spectral image of the vHS1 at 11 K, 140 K, and 260 K. The vHS point visibly shifts in energy as temperature is varied. To examine the trend systematically, we extract the energy position of vHS for each temperature. In Fig. 3a, we identify the bands leading to vHS1. Since the vHS point is slightly above $E_F$, to track its energy position, we first fit the band dispersion at each temperature from a momentum distribution curve (MDC) analysis. From the MDC, we identify a main peak next to another feature in the form of a smaller peak. We fit the MDC line profile to two Lorentzian functions on a Gaussian background. The location of the vHS band can therefore be extracted as a function of binding energy (Fig. 3d). We then fit for the location of the band top using a parabolic function of the dispersion at each temperature. We note that only the dispersion beyond -30 meV is used in the parabolic fit to avoid any complication due to the potential opening of a CDW gap or dispersion kink. The extracted location of vHS1 is plotted as a function of temperature in Fig. 3i, showing an upward shift as temperature is lowered towards $T_{CDW}$. Additional evidence for the band shift is seen in the peak shift from both energy distribution curves (EDCs) and MDCs as a function of temperature (see Figure S7). The trend of the shift is consistent with the ordering of the magnetic moment ($m$), as can be seen by overlaying in Fig. 3h the $m^2$ measured from neutron diffraction[20].

As the vHSs are shifted to the vicinity of $E_F$ by the magnetic order, we therefore examine the potential role of the vHSs in the formation of the CDW order. At 140 K above the onset of the CDW order, vHS1 is above $E_F$ while vHS2 is below $E_F$. They form distinct Fermi surfaces. In particular, the Fermi surface from vHS1 ($xy/x^2$-$y^2$) can be better observed under an in-plane polarization (linear vertical, LV) while vHS2 ($xz/yz$) can be better observed in the linear horizontal (LH) polarization that contains an out-of-plane component (Fig. 4a). Noticeably, the vHS1 band forms a circular Fermi surface while that of vHS2 band resembles more that of the triangular Fermi surface predicted by Kagome models at vHS filling[4]. We can further confirm their contribution to potential nesting condition using DFT-based Lindhard susceptibility calculations (see Figure S11). While DFT does not accurately reproduce the Fermi surface, we can tune the chemical potential to identify energies where such Fermi surface contours are revealed. A susceptibility calculation at such energy that reproduces the vHS2 Fermiology indeed produces a peak at the 2x2 CDW wavevector while that done for the vHS1 Fermiology does not (see Figure S11). This suggests that the vHS contribution to the CDW formation may be orbital-dependent in FeGe, as has also been reported for $CsV_3Sb_5$[45].

Furthermore, we can identify participation of the vHS bands in the CDW ordering by comparing the band dispersions across $T_{CDW}$. From a comparison of the Fermi surface at 140 K and 13 K (Figs. 4a, 4b), we see that the vHS2 undergoes the largest change. This can also be seen by comparing two cuts across $T_{CDW}$. On the cut along $\bar{K}$-$\bar{M}$-$\bar{K}$ (Figs. 4c and 4d), even though matrix elements suppress the spectral weight of vHS2, we can still identify the presence of $k_z$-broadened vHS2 via a hump in the EDC (Fig. 4g). In the CDW phase, this band appears to be split into a lower branch and an additional band with small velocity across $E_F$, seen as a split peak in the EDC (Figs. 4d and 4g). Such band splitting at the M point is very similar to that observed in CsV$_3$Sb$_5$[45], and is interpreted to be the result of folding and gap opening at the M point. On a parallel cut away from $\bar{K}$-$\bar{M}$-$\bar{K}$, gap opening can also be observed on vHS2 with a typical bend-back behavior of the dispersion (Figs. 4e, 4f, 4h, and Figure S9). Besides vHS2, we also observe spectral gap opening on vHS1. Spectral images across the $\bar{K}$-$\bar{M}$-$\bar{K}$ cut are shown for 140 K > $T_{CDW}$ (left side of Fig. 4i) and 70 K < $T_{CDW}$ (right side of Figs. 4i). The spectra have been divided by the Fermi-Dirac function convolved with the instrument resolution to also uncover spectral weight above $E_F$. Noticeably, a spectral weight suppression is observed on vHS1, not on any other bands crossing $E_F$. This is further confirmed by the EDC for the vHS1 at $k_{vHS}$, where a spectral weight suppression below $T_{CDW}$ indicates the gap opening (Fig. 4j and Figure S8). The observable lower edge of the gap below $E_F$ is -20 meV, consistent with the observed full gap size of 50 meV by scanning tunneling microscopy[37]. In contrast, the EDCs for the other band crossings away from the BZ edge L points in the CDW ordered phase show ungapped peaks at $E_F$ in the CDW phase, indicating that the CDW gap is momentum-dependent. In Fig. 4k, we show a detailed temperature evolution of the integrated spectral weight within the gap energy of 20 meV of $E_F$. The smoothed average of the scattered data points (red line in Fig. 4k) shows a trend consistent with the onset of the depletion of the spectral weight at $T_{CDW}$ = 100 K (see Figure S8).

Lastly, we report on the observation of electron-phonon coupling in FeGe. Dispersion kinks have been reported in unconventional superconductors such as the cuprates and iron-based superconductors[17-23,44,46,47]. The origin of such dispersion kinks has been proposed to indicate electron-boson coupling, to either phonons or magnons. On the vHS1 band where a CDW gap is observed, a kink structure is observed in the electronic dispersion (Fig. 5a). We can use MDC analysis to extract the real and imaginary parts of the self-energy from the peak position and the full width at half-max of the MDC fitting, respectively (Figs. 5b-d), showing the coupling of the electrons to a bosonic mode at ~30 meV below $E_F$. We can then extract an electron-boson coupling constant from the slope of the real part of the self-energy as well as from the ratio of the renormalized band velocity ($v_F$) and the bare band velocity ($v_{F^0}$), both of which give a consistent coupling constant $\lambda$ of 0.5~0.6[48,49]. Notably, this kink structure has little resolvable temperature dependence from 11 K ($T < T_{canting}$) to 70 K ($T_{canting} < T < T_{CDW}$) to 140 K ($T_{CDW} < T$).

To identify the origin of this bosonic mode, we carried out INS measurement. As presented in Fig. 5f, the observed spin waves stemming from AFM zone center and in-plane Γ point do not cross the M point near 30 meV, thus ruling out direct electron-magnon interaction at the M point. In contrast, the phonon spectra do reveal optical modes near 33~38 meV around the M point (Fig. 5e). More interestingly, a hardening of this optical phonon mode at 33 meV is observed across the $T_{CDW}$ from 120 K to 70 K, appearing at the M point and extending to the K points (Fig. 5i and Figure S10b), but absent at the Γ points (Fig. 5h). This optical phonon mode, therefore, is likely the mode responsible for the dispersion kink that we observe from ARPES. We note that the size of the hardening is about 1 meV, which is too small to be resolved in the temperature-dependent kink analysis (Fig. 5a). To identify the phonon mode, we carried out phonon calculations using

DFT, which match excellently with the measured phonon spectra (Fig. 5e). The optical phonon is determined to be an $A_{2u}$ mode involving the movement of the Fe atoms from the Kagome plane as well as movement of the inter-layer Ge atoms (Fig. 5j). In addition, we observe no acoustic phonon softening from 8 K to 70 K (see Figure S10a), contrary to the hardening in the $A_{2u}$ optical mode across the same temperature region. The electron-phonon coupling here in FeGe is similar in many aspects to the $AV_3Sb_5$ system, where a dispersion kink at a similar energy scale of 36 meV has also been reported[50]. In addition, a hardening of the $B_{3u}$ longitudinal optical phonon mode around 10 meV has been observed in $AV_3Sb_5$ across its $T_{CDW}$[51] while no acoustic phonon softening is observed[50,52-55].

Having presented all the data, we now discuss the implications. We have experimentally identified potential key ingredients for the formation of the CDW order in magnetic FeGe, namely the orbital-dependent vHSs that are brought to the vicinity of $E_F$ due to the magnetic splitting of the bands. The contribution of the vHSs to the formation of the CDW order is evident in the opening of gaps at $E_F$ in the CDW phase while other bands remain gapless. In addition, we identify electron-phonon coupling around the M point of the BZ that is connected by the CDW wavevector. We can first compare the phenomenology of the CDW order in FeGe to that in non-magnetic kagome $AV_3Sb_5$[17-19,45,50,56,57]. In both CDW ordered phases, vHSs are observed to be in the vicinity of $E_F$, albeit it is accidental in $AV_3Sb_5$ and magnetism-assisted in FeGe. The vHSs are observed to have orbital-dependent contributions gauged from their fermiology. In both cases, the *xz/yz* orbitals are proposed to be dominant[56]. The CDW gap at $E_F$ is observed at the BZ edge, which is connected by wavevectors that correspond to the diffraction peaks seen by x-ray and neutrons. Electron-phonon coupling is present in both systems in the absence of acoustic phonon softening at the CDW wavevectors[50,52-55]. These similarities point to a potential universality of the phenomenology of CDW amongst the kagome metals. However, the microscopic origin of the CDW order in kagome system is still unresolved. In the weakly-correlated AV$_3$Sb$_5$ system, nesting of vHSs has been proposed based on 2D models. In FeGe, however, the moderate electron correlations deem a purely nesting-driven scenario unlikely. This is reflected in the larger degree of mismatch between DFT and measured dispersions in FeGe as well as difficulty to stabilize a CDW order from DFT alone. If nesting is indeed the dominant driver, then it may be expected that other A-type AFM Kagome system, such as FeSn[11], could also develop 2x2 CDW order when the spin splitting shifts the vHSs at M into the proximity of $E_F$. Since FeGe is the only kagome lattice magnet with CDW discovered so far, it is therefore likely that correlation effects, perhaps even orbital-dependent correlation effects, must be taken into consideration in addition to the vHS presence near $E_F$ to understand the origin of the CDW order in magnetic kagome systems. Regardless of the ultimate microscopic theoretical description of the CDW on stacked kagome lattices, our work reveals an intimate interaction of the CDW order and magnetism in a moderately correlated kagome metal and provides an experimental groundwork towards a potential universal understanding of CDW order in kagome systems.

**Methods:**

Single crystals were synthesized using chemical vapor transport[20]. ARPES measurements were carried out at BL 5-2 of the Stanford Synchrotron Radiation Lightsource and the MAESTRO beamline (7.0.2) of the Advanced Light Source, with a DA30 electron analyzer and a R4000 electron analyzer with deflector mode, respectively. FeGe single crystals were cleaved *in-situ* in ultra-high vacuum with a base pressure better than $5 \times 10^{-11}$ Torr. Samples were cleaved in the

(001) orientation. Two types of terminations were found, one corresponding to the kagome layer and another to the Ge layer. Both bulk and surface states were identified on both types of terminations. A detailed discussion on the terminations is provided in SI. The data shown in the main text come from the Ge termination unless specifically stated otherwise, but all the features discussed are those that are found in both terminations, and hence representative of the bulk kagome lattice. Energy and angular resolutions used were better than 20 meV and 0.1°, respectively. Measurements shown were taken with a beam spot smaller than 50 μm × 50 μm. All gap measurements were done with correction of the Fermi level by a measurement of a polycrystalline gold that was electrically connected to the sample.

INS experiments were carried out on ARCS spectrometer, Spallation Neutron Source, Oak Ridge National Laboratory (ORNL)[58] on 0.9 grams of co-aligned single-crystal samples in the [$H, H, L$] scattering plane. We measured the neutron spectrum at 120 K, 70 K and 8 K using Horace mode[59] with 120 meV (Fig. 5f) and 45 meV (Fig. 5e) incident energy. Magnon and phonon spectrum are plotted using low-$Q$ ($L$ = [0, 2.6]) and high-$Q$ ($L$ = [2.9, 3.1]) data, respectively. The vibration direction of phonon observed by INS is always parallel to the momentum transfer $Q$ (Fig. 5g), therefore the phonon mode shown in Fig. 5e is mostly the $A_{2u}$ mode vibrating along the $c$-axis. Here we do constant-$Q$ cuts at the Γ and M points (Figs. 5h and 5i, respectively), then fit the measured phonon data using Lorentzian subtracted by a linear background.

DFT calculations are performed with the Vienna *ab-initio* simulation package[60,61]. The exchange-correlation interaction between electrons is mimicked with generalized gradient approximation as parameterized by Perdew-Burke-Ernzerhof[62]. A cutoff energy of 350 eV and a $k$-mesh of 12×12×8 are used throughout. Experimental lattice constants are employed, and the atomic positions are fully relaxed until the remaining force is less than 1 meV/Å. Both the paramagnetic and antiferromagnetic structures are calculated. All calculations are performed considering the spin-orbital coupling, except for the structure relaxation and phonon dispersion calculation. The phonon dispersion is calculated by the finite displacement method as implemented in the PHONONPY software[63]. To calculate the Fermi surface and electron Lindhard susceptibility, we first fit the DFT band structure with wannier orbitals (Fe $d$ and Ge $p$) as implemented in Wannier90 software[64] and then calculate the three-dimensional Fermi surface in the full Brillouin zone based on the obtained tight-binding Hamiltonian with a k-mesh of 150×150×80. The electron susceptibility is calculated based on the Fermi surface by employing the method described in Ref. [65].

**Data availability.** The data that support the plots within this paper and other finding of this study are available from the corresponding authors upon reasonable request.

**Supplementary Information** is linked to the online version of the paper at xxx.

**Acknowledgements**

We thank Jianxin Zhu, Christopher Lane, Qimiao Si, Chandan Setty for helpful discussions. The ARPES work is supported by the Gordon and Betty Moore Foundation's EPiQS Initiative through grant no. GBMF9470, the Robert A. Welch Foundation Grant No. C-2024, and the U.S. Department of Energy (DOE) grant No. DE-SC0021421. The neutron scattering and single crystal synthesis work at Rice was supported by US NSF-DMR-2100741 and by the Robert A. Welch Foundation under Grant No. C-1839, respectively. The work at the University of California,

Berkeley was supported by the U.S. DOE under Contract No. DE-AC02-05-CH11231 within the Quantum Materials Program (KC2202). This research used resources of the Advanced Light Source and the Stanford Synchrotron Radiation Lightsource, both U.S. DOE Office of Science User Facilities under contract Nos. DE-AC02-05CH11231 and AC02-76SF00515, respectively. A portion of this research used resources at the Spallation Neutron Source, a DOE Office of Science User Facility operated by ORNL. B.Y. acknowledges the financial support by the European Research Council (ERC Consolidator Grant ``NonlinearTopo'', No. 815869) and the ISF - Personal Research Grant (No. 2932/21).

**Author Contributions**

M.Y., P.D., and R.J.B. managed the project. The single crystal FeGe samples were grown by X.T. and B.G. under the guidance of P.D.. APRES experiments were carried out by J.S.O., X.T., J.H. and M.Y. with the assistance of M.H., D.L., C.J., A.B., and E.R.. First-principle calculations were performed by H.T. and B.Y.. Neutron scattering measurements and analysis were carried out by X.T., L.C., and P.D.. The paper was written by M.Y., P.D., X.T., J.S.O., and L.C. with inputs and significant discussions from all coauthors.

The authors declare no competing financial interests.

Correspondence and requests for materials should be addressed to M.Y. (mingyi@rice.edu) or P.D. (pdai@rice.edu) or R.J.B. (robertjb@berkeley.edu).

**References**

[1] Paschen, S. and Si, Q. Quantum phases driven by strong correlations. *Nat. Rev. Phys.* **3**, 9-26 (2021).
[2] Syôzi, I. Statistics of kagomé lattice. *Prog. Theor. Phys.* **6**, 306–308 (1951).
[3] Mazin, I. I. *et al.* Theoretical prediction of a strongly correlated Dirac metal. *Nat. Commun.* **5**, 4261, (2014).
[4] Kiesel, M. L., Platt, C., Thomale, R. Unconventional Fermi Surface Instabilities in the Kagome Hubbard Model. *Phys. Rev. Lett.* **110**, 126405 (2013).
[5] Mielke, A. Ferromagnetic ground states for the Hubbard model on line graphs. *J. Phys. A* **24**, L73 (1991).
[6] Tanaka, A. and Ueda, H. Stability of Ferromagnetism in the Hubbard Model on the Kagome Lattice. *Phys. Rev. Lett.* **90**, 067204 (2003).
[7] Kuroda, K. *et al*. Evidence for magnetic Weyl fermions in a correlated metal. *Nat. Mater.* **16**, 1090-1095 (2017).
[8] Lin, Z. *et al*. Flatbands and Emergent Ferromagnetic Ordering in $Fe_3Sn_2$ Kagome Lattices. *Phys. Rev. Lett.* **121**, 096401 (2018).
[9] Ye, L. *et al.* Massive Dirac fermions in a ferromagnetic kagome metal. *Nature* **555**, 638-642, (2018).
[10] Kang, M. *et al*. Dirac fermions and flat bands in the ideal kagome metal FeSn. *Nat. Mater.* **19**, 163-169 (2020).
[11] Xie, Y. *et al.* Spin excitations in metallic kagome lattice FeSn and CoSn. *Commun. Phys.* **4**, 240 (2021).
[12] Yin, J.-X. *et al*. Negative flat band magnetism in a spin-orbit-coupled correlated kagome magnet. *Nat. Phys.* **15**, 443-448 (2019).
[13] Yin, J.-X. *et al*. Quantum-limit Chern topological magnetism in $TbMn_6Sn_6$. *Nature* **583**, 533-536 (2020).
[14] Li, M. *et al*. Dirac cone, flat band and saddle point in kagome magnet $YMn_6Sn_6$. *Nat. Commun.* **12**, 3129 (2021).
[15] Morali, Noam, *et al*. Fermi-arc diversity on surface terminations of the magnetic Weyl semimetal $Co_3Sn_2S_2$. *Science* **365**,1286-1291 (2019).
[16] Liu, D. F., *et al*. Magnetic Weyl semimetal phase in a Kagomé crystal. *Science* **365**, 1282-1285 (2019).
[17] Chen, Hui, *et al*. Roton pair density wave in a strong-coupling kagome superconductor. *Nature* **599**, 222-228 (2021).
[18] Ortiz, B. R. *et al*. $CsV_3Sb_5$: A $Z_2$ Topological Kagome Metal with a Superconducting Ground State. *Phys. Rev. Lett.* **125,** 247002 (2020).
[19] Jiang, K. *et al.* Kagome superconductors $AV_3Sb_5$ (A=K, Rb, Cs). Preprint at https://arxiv.org/abs/2109.10809 (2021).
[20] Teng, X. *et al*. Discovery of charge density wave in a correlated kagome lattice antiferromagnet. Nature **609**, 490 (2022).
[21] Keimer, B. *et al*. From quantum matter to high-temperature superconductivity in copper oxides. *Nature* **518**, 179-186 (2015).
[22] Si, Q., Yu, R., Abrahams, E. High-temperature superconductivity in iron pnictides and chalcogenides. *Nat. Rev. Mater.* **1**, 16017 (2016).
[23] Fernandes, R. M. *et al*. Iron pnictides and chalcogenides: a new paradigm for superconductivity. *Nature* **601**, 35-44 (2022).

[24] Yi, M., Zhang, Y., Shen, Z.-X. Lu, D. Role of the orbital degree of freedom in iron-based superconductors. *npj Quantum Mater*. **2**, 57 (2017).
[25] Zhou, Y., Kanoda, K., Ng, T.-K. Quantum spin liquid states. *Rev. Mod. Phys.* **89**, 025003 (2017).
[26] Wang, Q. *et al.* Field-induced topological Hall effect and double-fan spin structure with a c-axis component in the metallic kagome antiferromagnetic compound $YMn_6Sn_6$. *Phys. Rev. B* **103**, 014416 (2021).
[27] Jiang, Y.-X. et al. Unconventional chiral charge order in kagome superconductor $KV_3Sb_5$. *Nat. Mater.* **20**, 1353–1357 (2021).
[28] Mielke, C. *et al.* Time-reversal symmetry-breaking charge order in a correlated kagome superconductor. *Nature* **602**, 245-250 (2022).
[29] Feng, X., Jiang, K., Wang, Z. & Hu, J. Chiral flux phase in the Kagome superconductor $AV_3Sb_5$. *Sci. Bull.* **66**, 1384–1388 (2021).
[30] Denner, M. M., Thomale, R. & Neupert, T. Analysis of charge order in the kagome metal $AV_3Sb_5$ (A = K, Rb, Cs). *Phys. Rev. Lett.* **127**, 217601 (2021).
[31] Lin, Y.-P. & Nandkishore, R. M. Complex charge density waves at Van Hove singularity on hexagonal lattices: Haldane-model phase diagram and potential realization in the kagome metals $AV_3Sb_5$ (A=K, Rb, Cs). *Phys. Rev. B* **104**, 045122 (2021).
[32] Park, T., Ye, M. & Balents, L. Electronic instabilities of kagome metals: Saddle points and Landau theory. *Phys. Rev. B* **104**, 035142 (2021).
[33] Setty, C., Hu, H. Chen, L., Si, Q. Electron correlations and -breaking density wave order in a kagome metal. arXiv:2105.15204.
[34] Sales, B. C. *et al*. Tuning the flat bands of the kagome metal CoSn with Fe, In, or Ni doping. *Phys. Rev. Mater.* **5**, 044202 (2021).
[35] Li, H. *et al*. Conjoined Charge Density Waves in the Kagome Superconductor $CV_3Sb_5$. Preprint at https://arxiv.org/abs/2202.13530 (2022).
[36] Qian, T. *et al*. Revealing the competition between charge density wave and superconductivity in $CsV_3Sb_5$ through uniaxial strain. *Phys. Rev. B* **104**, 144506 (2021).
[37] Yin, J.-X. *et al*. Discovery of charge order and corresponding edge state in kagome magnet FeGe. Phys. Rev. Lett. **129**, 166401 (2022).
[38] Ohoyama, T., Kanematsu, K. & Yasukōchi, K. A New Intermetallic Compound FeGe. *J. Phys. Soc. Jpn.* **18**, 589-589 (1963).
[39] Bernhard, J., Lebech, B. & Beckman, O. Neutron diffraction studies of the low-temperature magnetic structure of hexagonal FeGe. *J. Phys. F: Met. Phys.* **14**, 2379-2393 (1984).
[40] Bernhard, J., Lebech, B. & Beckman, O. Magnetic phase diagram of hexagonal FeGe determined by neutron diffraction. *J. Phys. F: Met. Phys.* **18**, 539-552 (1988).
[41] Huang, L. & Lu, H. Signatures of Hundness in kagome metals. *Phys. Rev. B* **102**, 125130 (2020).
[42] Setty, C. *et al*. Electron correlations and charge density wave in the topological kagome metal FeGe. Preprint at https://arxiv.org/abs/2203.01930 (2022).
[43] Yi, M. *et al*. Role of the orbital degree of freedom in the iron-based superconductors. *npj Quantum Materials* **2**, 57 (2017).
[44] Sobota, J., He, Y., Shen, Z.-X. Angle-resolved photoemission studies of quantum materials. *Rev. Mod. Phys.* **93** 025006 (2021).

[45] Kang, M. *et al.* Twofold van Hove singularity and origin of charge order in topological kagome superconductor $CsV_3Sb_5$. *Nat. Phys.* **18**, 301-308 (2021).
[46] Yu, T.L. *et al.* Colossal Band Renormalization and Stoner Ferromagnetism induce by electron-antiferromagnetic-magnon coupling. Preprint at https://arxiv.org/abs/2011.05683 (2020).
[47] Wray, L. *et al.* Momentum dependence of superconducting gap, strong-coupling dispersion kink, and tightly bound Cooper pairs in the high-Tc $(Sr,Ba)_{1-x}(K,Na)_xFe_2As_2$ superconductors. *Phys. Rev. B* **78**, 184508 (2008).
[48] Göran Grimvall, The Electron-Phonon Interaction in Metals, edited by E. Wohlfarth (North-Holland Pub. Co., New York, 1981).
[49] E.W. Plummer, J.R. Shi, S.J. Tang, E. Rotenberg, and S.D. Kevan, Enhanced electron-phonon coupling at metal surfaces. *Prog. Surf. Sci.* **74**, 251 (2003).
[50] Luo, H. *et al.* Electronic nature of charge density wave and electron-phonon coupling in kagome superconductor $KV_3Sb_5$. *Nat. Commun.* **13**, 273 (2022).
[51] Xie, Y. *et al.* Electron-phonon coupling in the charge density wave state of $CsV_3Sb_5$. *Phys. Rev. B* **105**, L140501 (2022).
[52] Tan, Hengxin *et al.* Charge Density Waves and Electronic Properties of Superconducting Kagome Metals. *Phys. Rev. Lett.* **127**, 046401 (2021).
[53] Li, Haoxiang et al. Observation of Unconventional Charge Density Wave without Acoustic Phonon Anomaly in Kagome Superconductors $AV_3Sb_5$ (A =Rb, Cs). *Phys. Rev. X* **11**, 031050 (2021).
[54] Wu, Shangfei *et al.* Charge density wave order in kagome metal $AV_3Sb_5$ (A= Cs, Rb, K). Preprint at https://arxiv.org/abs/2201.05188 (2022).
[55] Wang, Chongze *et al.* Origin of charge density wave in the layered kagome metal $CsV_3Sb_5$. *Phys. Rev. B* **105**, 045135 (2022).
[56] Liu, Z. *et al.* Charge-Density-Wave-Induced Bands Renormalization and Energy Gaps in a Kagome Superconductor $RbV_3Sb_5$. *Phys. Rev. X* **11**, 041010 (2021).
[57] Nakayama, K. *et al.* Multiple energy scales and anisotropic energy gap in the charge-density-wave phase of the kagome superconductor $CsV_3Sb_5$. *Phys. Rev. B* **104**, L161112 (2021).
[58] Abernathy, D. L. *et al.* Design and operation of the wide angular-range chopper spectrometer ARCS at the Spallation Neutron Source. *Rev. Sci. Instrum* **83**, 15114 (2012).
[59] Ewings, R. A. *et al.* Horace: software for the analysis of data from single crystal spectroscopy experiments at time-of-flight neutron instruments. *Nucl. Instrum. Methods Phys. Res. A* **834**, 132 (2016).
[60] Kresse, Georg, and Jürgen Furthmüller. Efficient iterative schemes for ab initio total-energy calculations using a plane-wave basis set. *Phys. Rev. B* **54**, 11169 (1996).
[61] Kresse, Georg, and Jürgen Furthmüller. Efficiency of ab-initio total energy calculations for metals and semiconductors using a plane-wave basis set. *Comput. Mater. Sci.* **6**, 15-50 (1996).
[62] Perdew, John P., Kieron Burke, and Matthias Ernzerhof. Generalized gradient approximation made simple. *Phys. Rev. L* **77**, 3865 (1996).
[63] Togo, Atsushi, and Isao Tanaka. First principles phonon calculations in materials science. *Scr. Mater.* **108**, 1 (2015).
[64] Mostofi, Arash A., *et al.* "wannier90: A tool for obtaining maximally-localised Wannier functions." *Comput. Phys. Commun.* **178**, 685 (2008).

[65]   Johannes, M. D., and I. I. Mazin. Fermi surface nesting and the origin of charge density waves in metals. *Phys. Rev. B* **77**, 165135 (2008)

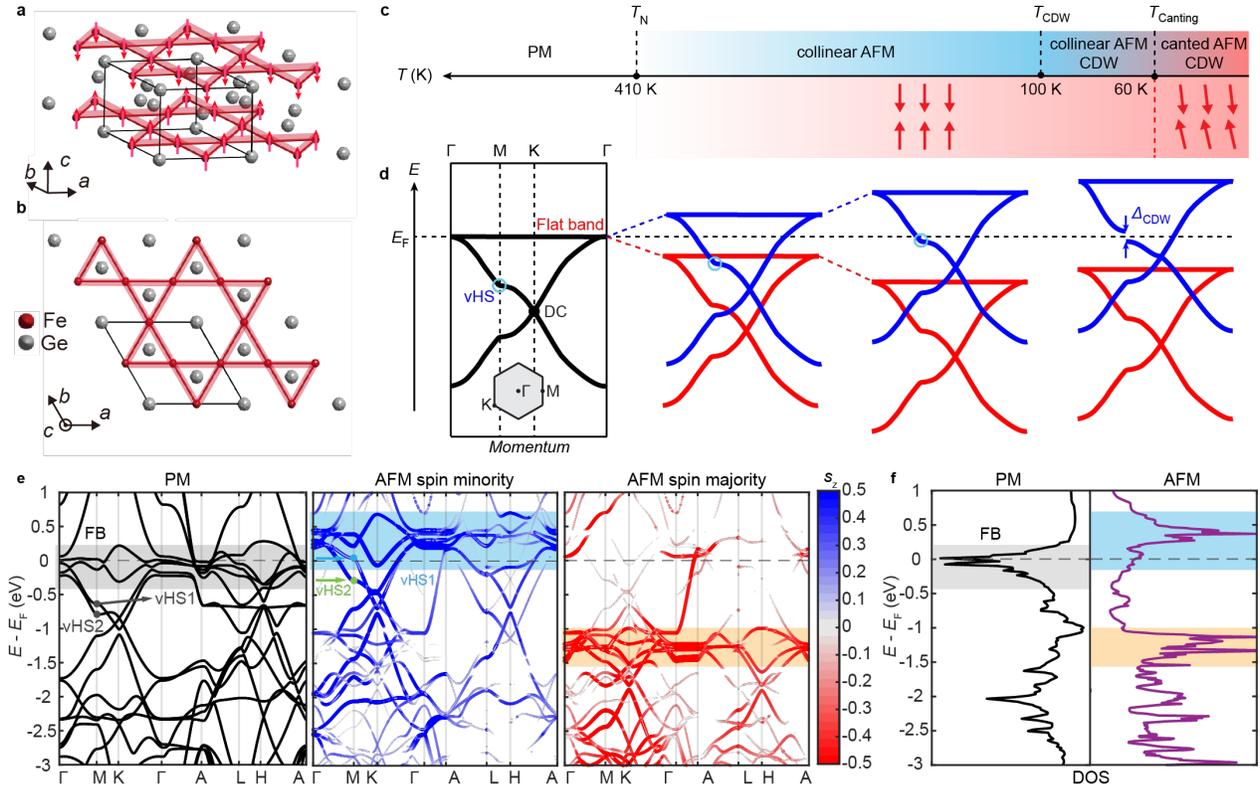

**Figure 1. Crystal structure and electronic structure.** (a) Crystal structure of hexagonal FeGe with unit cell shown by solid lines. A-type AFM order is illustrated by red arrows denoting spin ordering. (b) A top view of the crystal structure to visualize the Fe kagome layer. (c) Summary of the transitions as a function of temperature. (d) Simplified schematic summary of AFM-induced band splitting. In the PM state, the kagome flat band appears at $E_F$, with a Dirac crossing and a pair of vHSs well below $E_F$. In the AFM phase, the ordered moment within each ferromagnetic layer induces exchange splitting of the spin-minority and spin-majority bands, resulting in a upshift of the blue vHS towards $E_F$. At the onset of the CDW order, a gap opens at the vHS band near $E_F$. The temperature evolution (horizontal direction) is aligned with the temperature axis in (c). (e) DFT calculated band structures for the PM state, as well as those for the spin minority and majority bands within a FM layer in the AFM state. Flat band regions are highlighted with colored shades. (f) DOS calculations for the PM and AFM phases.

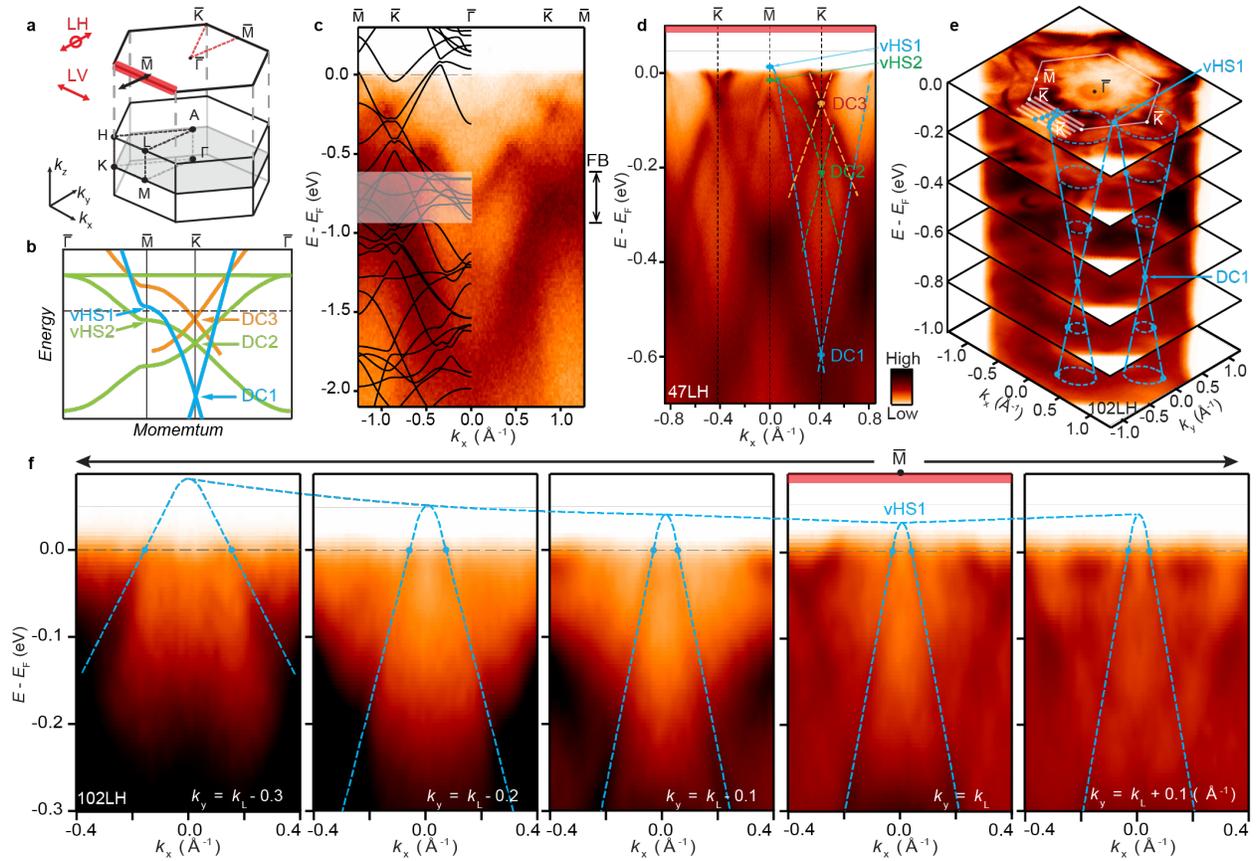

**Figure 2. Key signatures of kagome band structure.** (a) 3D and 2D projected BZ notations with polarization vectors indicated. (b) Schematic of DCs and associated vHSs. (c) Large energy range dispersions measured compared to that from DFT calculation after renormalization by a factor of 1.6. The lower kagome flat bands are highlighted similar to the way of Fig. 1e. (d) Spectral image measured along $\bar{\text{K}}$-$\bar{\text{M}}$-$\bar{\text{K}}$ using LH 47 eV photons. (e) Constant energy contours measured with 102 eV photons and LH polarization. Blue dotted lines indicate a Dirac cone at the $\bar{\text{K}}$ point and the corresponding vHS at $\bar{\text{M}}$. (f) Stack of cuts taken along $k_x$ at $k_y = k_L - 0.3$, $k_L - 0.2$, $k_L - 0.1$, $k_L$, $k_L + 0.1$ (Å$^{-1}$), as indicated in (e) as white solid lines. The Fermi crossings are marked by blue dots, also marked on the corresponding cuts in (d). All data are taken at 11 K.

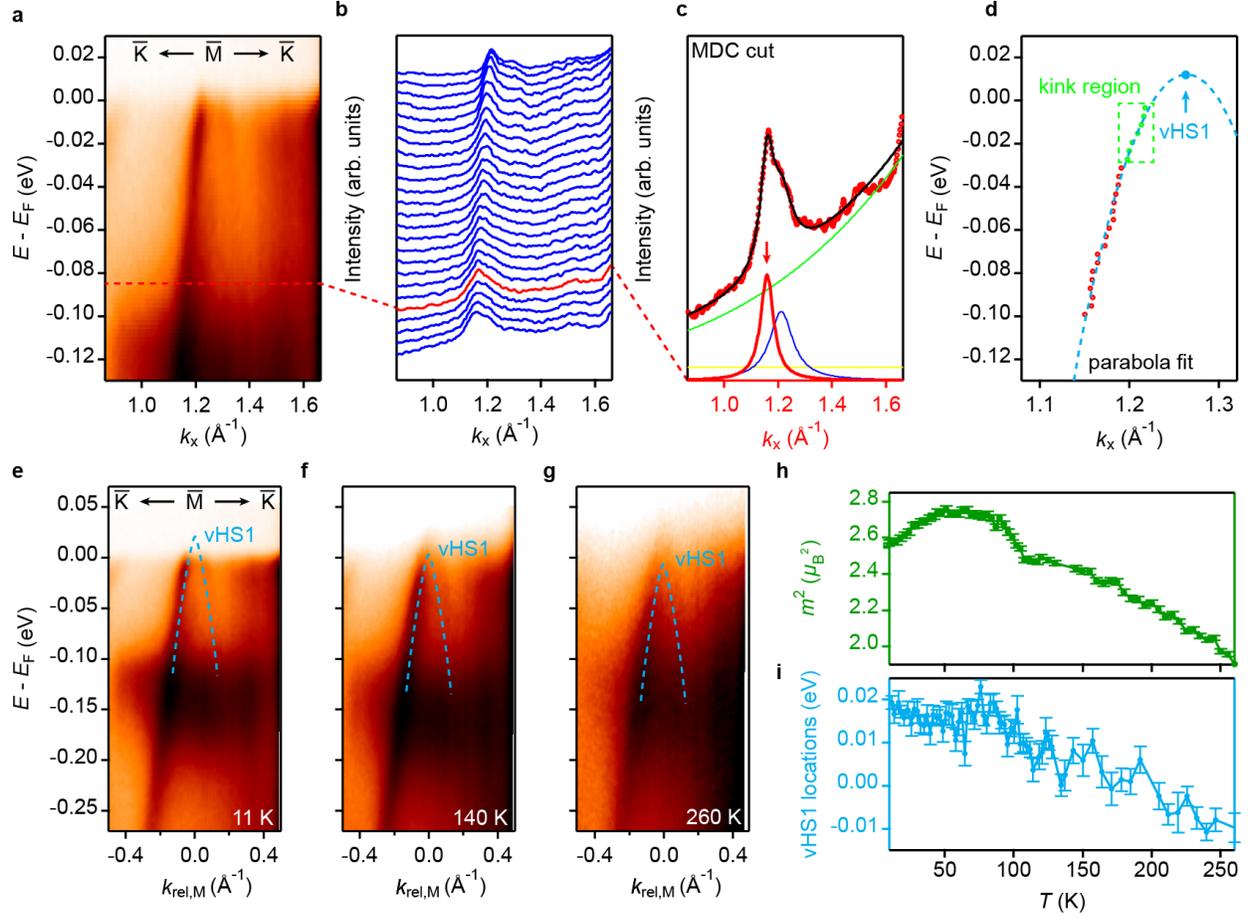

**Figure 3. Temperature evolution of van Hove singularity.** (a) Spectral images across the M point along $\bar{K}$-$\bar{M}$-$\bar{K}$, taken with LH-polarized 69 eV photons (b) A stack of MDCs (from $E_F$ to -0.1 eV) from (a). Red line in the stack is the MDC curve along the red dashed line in (a). (c) The MDC from the red dashed line in (a) and the red curve in (b). We fit the curve with two Lorentzians (red and blue), one background Gaussian profile (light green) and a constant background (yellow) curves. (d) Fitting results from the MDC stack. Red circles are peak positions of vHS1-forming band dispersions (red curve in (b)). We use a parabola fit to extrapolate locations of vHS1 above $E_F$. (e-g) Spectral images across the $\bar{M}$ point along $\bar{K}$-$\bar{M}$-$\bar{K}$, showing temperature evolution of locations of vHS1. (h) Ordered magnetic moment measured by neutron scattering reproduced from Ref. [20] (green) (i) Fitted vHS1 locations.

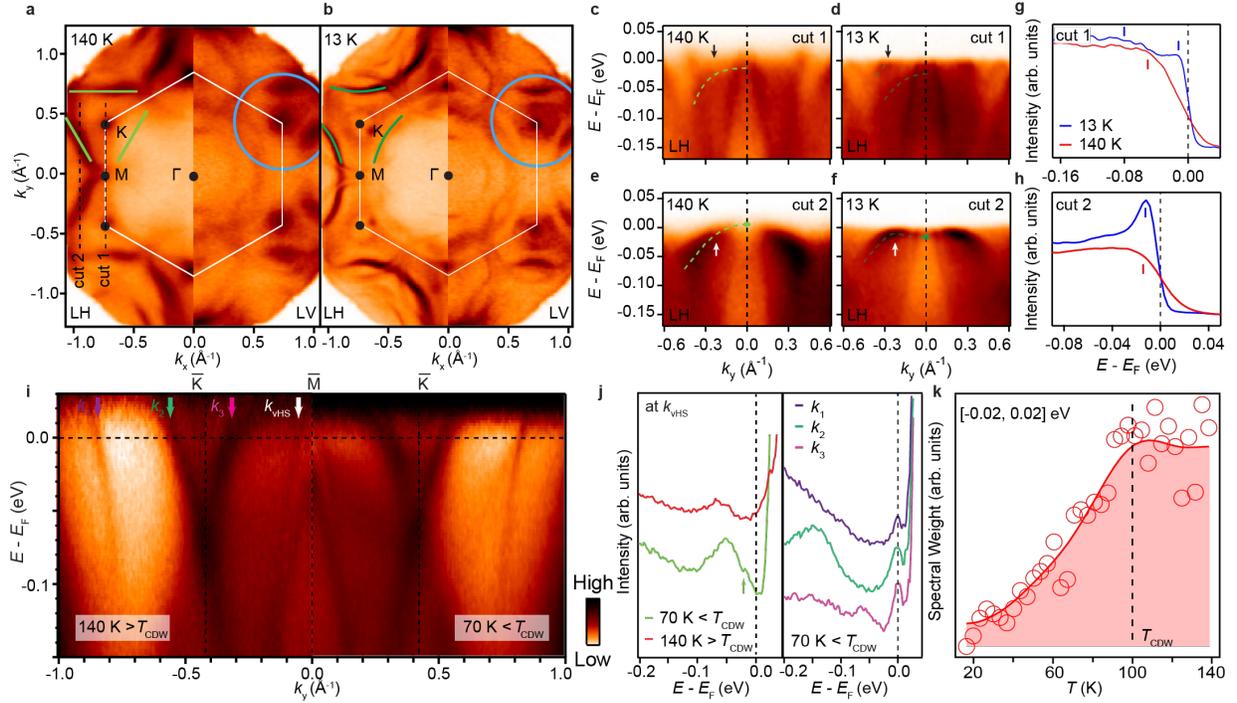

**Figure 4. Observation of CDW gap.** (a) Fermi surface taken on kagome termination at 140 K with 69 eV photons, with the left half under LH polarization and right half under LV polarization. (b) Same as (a) but taken at 13 K. (c)-(d) $\overline{K}$-$\overline{M}$-$\overline{K}$ cut (cut 1) from (a) and (b). (e)-(f) cut 2 from (a) and (b) showing CDW gap opening on vHS2 band. (g) EDC from cut 1 marked by the arrow showing the appearance of two features at 13 K. (h) EDC from cut 2 marked by the arrow showing the gap opening. (i) Spectral image taken with LH-polarized 47 eV photons across the $\overline{K}$-$\overline{M}$-$\overline{K}$ high symmetry direction at 140 K > $T_{CDW}$ (left) and 70 K < $T_{CDW}$ (right). The data has been divided by the Fermi Dirac function convolved with the energy resolution to reveal signal above $E_F$. The Fermi momenta of four band dispersions are indicated ($k_1$, $k_2$, $k_3$, $k_{vHS}$) where $k_{vHS}$ corresponds to the vHS crossing. (j) EDC at $k_{vHS}$, showing a gap below $T_{CDW}$, as well as EDCs for other band crossings as labeled in (i) showing no gap opening. (k) Integrated spectral weight across the CDW gap energy between [-0.02, 0.02] eV at $k_{vHS}$ as a function of temperature (red circles). The red line is the smoothed average of the data points, showing a suppression onset at $T_{CDW}$.

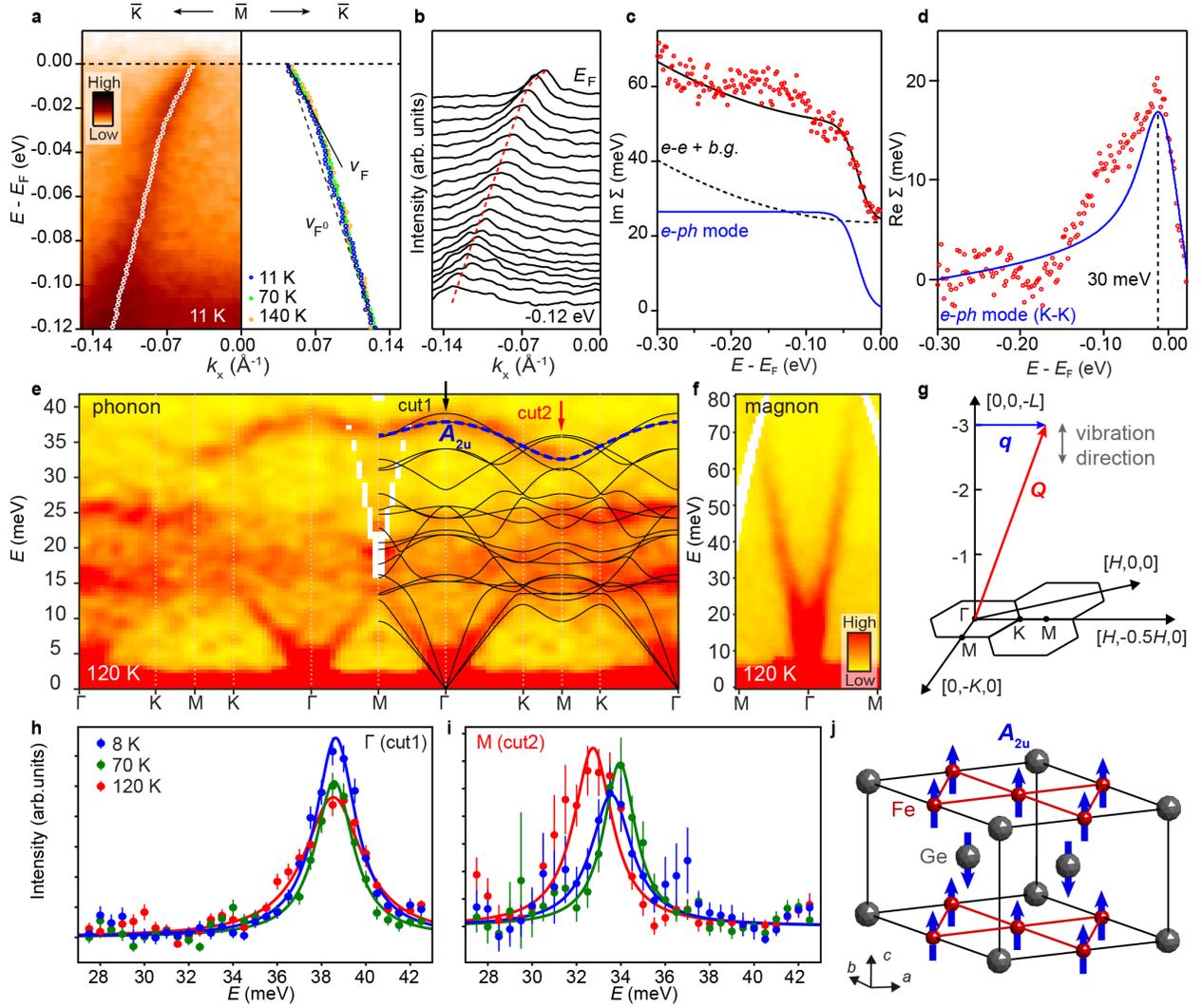

**Figure 5. Electron-boson coupling in FeGe.** (a) vHS1-forming band dispersion (same one as that in Fig. 4). White dots are fitted band dispersions from MDC analysis. On the right side, we overlay MDC fitting results at 11 K ($T < T_{canting}$), 70 K ($T_{canting} < T < T_{CDW}$), and 140 K ($T_{CDW} < T$). (b) A stack of MDC curves directly visualizing the kink structure. (c) Imaginary part of self-energy derived from width analysis of MDC fittings. We introduce one electron-boson coupling mode together with background signals considering moderate electron-electron interaction (Fermi-liquid-like behavior). (d) Real part of self-energy analysis of peak position from MDC fittings. We overlay the Kramers-Kronig transformed electron-boson coupling mode from the imaginary part

of the self-energy (blue solid curve in (c)). A bosonic mode is observed at 30 meV. (e) INS measurement at 120 K of the in-plane phonon spectra at $L = [2.9, 3.1]$ overlaid with DFT calculated phonon spectra. (f) INS measurement of the in-plane magnon spectra with $L = [0, 2.6]$. (g) INS measurement geometry illustrating the relationship between neutron momentum transfer $Q$, phonon wavevector $q$, and vibrational direction. (h) Measured phonon spectra at the $\Gamma$ (0, 0, 3) point (cut 1), showing no temperature dependence. (i) Measured phonon spectra at the M points (cut 2, averaged between (1, -0.5, 3) and (-1, 0.5, 3)), showing hardening of the optical phonon mode across $T_{CDW}$. The fitted energy at (8 K, 70 K, 120 K) is (32.74±0.13, 33.94±0.10, 33.57±0.15) meV. (j) Identified phonon mode ($A_{2u}$) for the optical branch outlined in (e).